\documentclass{aa}

\usepackage{savesym}
\usepackage{amsmath}
\savesymbol{iint}
\savesymbol{iiint}
\usepackage{txfonts}
\restoresymbol{TXF}{iint}
\restoresymbol{TXF}{iiint}

\usepackage[version=3]{mhchem} 
\usepackage{color}
\usepackage[english]{babel}
\usepackage{rotating}
\usepackage{array}

\title{From interstellar carbon monosulfide to methyl mercaptan: paths of least resistance}
\author{T. Lamberts\inst{1,2}}

\institute{Institute for Theoretical Chemistry, University Stuttgart, Pfaffenwaldring 55, 70569 Stuttgart, Germany, \\Current Address: Leiden Institute of Chemistry, Gorlaeus Laboratories, Leiden University, P.O. Box 9502, 2300 RA Leiden, The Netherlands}

\date{Received XX/XX/XXXX / Accepted XX/XX/XXXX}

\abstract{ The 29 reactions linking carbon monosulfide (CS) to methyl mercaptan (\ce{CH3SH}) via ten intermediate radicals and molecules have been characterized with relevance to surface chemistry in cold interstellar ices. More intermediate species than previously considered are found likely to be present in these ices, such as trans- and cis-HCSH. Both activation and reaction energies have been calculated, along with low-temperature ($T > 45$~K) rate constants for the radical-neutral reactions. For barrierless radical-radical reactions on the other hand, branching ratios have been determined. The combination of these two sets of information provides, for the first time, quantitative information on the full \ce{H + CS} reaction network. Early on in this network, that is, early on in the lifetime of an interstellar cloud, HCS is the main radical, while later on this becomes first \ce{CH2SH} and finally \ce{CH3S}. }

\begin{document}


\maketitle

\section{Introduction}
In recent decades roughly 18 interstellar sulfurous compounds have been detected in the gas phase, such as carbon monosulfide (CS), thioformaldehyde (\ce{H2CS}), and methyl mercaptan (also known as methanethiol, \ce{CH3SH}) along with several of their isotopes \citep{Penzias:1971,Sinclair:1973,Liszt:1974,Linke:1979,Marcelino:2005,Mueller:2016}. All of these species have in fact been detected in interstellar regions related to star formation, such as the Orion A and Barnard 1 molecular clouds and hot molecular cores. Recently, exploiting the unprecedented sensitivity of ALMA within the PILS survey, \citet{Drozdovskaya:2018} showed that several of these S-bearing molecules in the outer disc-like structure of the protostar IRAS 16293–2422 B have been detected, and, moreover, they explored the chemical link with our solar system probed by the Rosetta mission on comet 67P/Churyumov-Gerasimenko \citep{Calmonte:2016}.

Additionally, \ce{CH3SH} has recently been proposed and modelled to be formed in the solid state via subsequent hydrogenation of CS \citep{Majumdar:2016,Mueller:2016,Vidal:2017}. The proposed surface formation stems from the assumed analogy between the \ce{H + CS} and the \ce{H + CO} reaction networks. Indeed methanol, \ce{CH3OH}, can be efficiently formed via surface hydrogenation reactions of CO \citep{Hiraoka:1998,Watanabe:2002,Fuchs:2009} and this pathway turns out to be important to reproduce interstellar methanol abundances \citep{Boogert:2015, Walsh:2016,Deokkeun:2017}.  {In fact, the first CS-containing molecule to be detected in the solid state in dense molecular clouds was OCS by \citet{Palumbo:1997}. They suggested this molecule to be present in a methanol-rich layer along with both \ce{H2CS} and \ce{CH3SH}.}  Within the reaction network starting from CS and leading up to the fully hydrogenated \ce{CH3SH}, not only are stable neutral species formed, but radicals such as HCS and \ce{CH2SH} can be created as well.  
These are subsequently available to react with other species in the ice, either by close proximity \citep{Fedoseev:2015} or when heavier radicals become mobile at slightly elevated
ice temperatures \citep{Garrod:2006a, Vasyunin:2013}. In this way, they may be incorporated into larger species, including those of astrobiological importance. It is known that two of the 21 aminoacids crucial for life contain -\ce{CH2SH} (Cysteine) and -\ce{SCH3} (Methionine) groups. The investigation of methyl mercaptan chemistry and the formation of intermediate reactive radicals thus directly provides a link to astrobiological studies.

Experimentally, there is a lack of laboratory data for surface reactions of \ce{H_nCS}-bearing species due to their chemical instability. The current research therefore provides a detailed theoretical investigation of the reaction network involving ten CS-bearing species and 29 reactions with atomic hydrogen,  explicitly taking quantum tunnelling into account. This work builds on various computational approaches previously reported in the literature where a part of the reaction network has been considered \citep{Kobayashi:2011, Kerr:2015, Vidal:2017}. To the best of our knowledge there are currently no low-temperature experimental data, calculated or measured unimolecular rate constants, or branching ratios available for any of the reactions mentioned below. Therefore, in this letter, for the first time, a full set of reactions within the hydrogen addition network is quantified, starting from carbon monosulfide and explicitly taking tunnelling into account, providing the astrochemical community with parameters that can be included in large-scale mean field models. 

\section{Methodology} 

Various levels of theory have been used in order to optimise the ratio between the computational cost and chemical accuracy. First of all, all calculations have been performed with density functional theory (DFT). Chemical accuracy for all calculated activation and reaction energies, as well as the rate constants is ensured by benchmarking the MPWB1K/def2-TZVP functional to a `higher' level of theory, namely CCSD(T)-F12/VTZ-F12. Furthermore, specifically and only for obtaining branching ratios for radical-radical reactions, the B3LYP/def2-TZVP and PBEh-3c/def2-mSVP combinations have been used. These relatively cheap functionals are expected to yield reliable results in terms of geometries, for example, to see which reactions take place. The computational details for both radical-neutral and radical-radical reactions are extensively discussed in Appendix~\ref{CompDet}.

\subsection{Surface model}
Reaction energetics, rate constants, and approximate branching ratios connecting the 12 species via 29 reactions are calculated with the use of density functional theory (DFT). 
Although \emph{surface} reactions of \ce{H_nCS} species with hydrogen atoms on interstellar ices are of interest here, for the current study, a { model was employed where the surface molecules are not explicitly taken into account.}

It was previously shown that the activation energy and rate constants are influenced by an environment of water ice molecules for the reaction \ce{H + H2O2} \citep{Lamberts:2016B,Lamberts:2017}. However, this is specifically a reaction with a flexible molecule that can form hydogen bonds. Indeed, this result cannot be generalized to be applicable to all species since we have also found that reactions with small molecules that are either less flexible or less capable of forming strong H-bonds may be much less affected by surrounding water molecules \citep{Meisner:2017, Lamberts:2017B}. In fact, for the methyl mercaptan molecule, the hydrogen bond strength is much less than for the O-substituted methanol \citep{Kosztolanyi:2003}. Also \citet{Du:2017} showed specifically that the activation energy of the abstraction reaction \ce{CH3SH + H -> CH3S + H2} is barely affected {with 0.1 kJ/mol} by the presence of a water molecule. Furthermore, \citet{Rimola:2014} have studied the influence of water ice on the reaction pathway starting from CO and leading to \ce{CH3OH} for a variety of binding modes on water clusters. They demonstrated that the activation energy of the key reactions is usually changed by $\sim$1 kJ/mol and in one case by 3.5 kJ/mol. 

As an additional check for the system at hand, we have calculated the activation energy, $E_\text{act.}$, of the first reaction of the reaction network, that is, \ce{H + CS -> HCS}, on two different water clusters of seven molecules. Both of the activation energies differ by less than 1 kJ/mol from the value in a pure gas-phase calculation; see Table~\ref{tabH+CS} and~\ref{ClusterTS}. 

\begin{table}[h]
 \centering
 \caption{Binding energies of CS and activation energies in kJ/mol of the reaction \ce{H + CS -> HCS} on two water clusters compared to gas-phase results.}\label{tabH+CS}
 \begin{tabular}{lll}
  & $E_\text{bind. CS}$ & $E_\text{act.}$ \\
  Gas phase & ... & 3.6\\
 Cluster 1 & -12.7 & 4.5 \\
 Cluster 2 & -11.0 & 3.2 \\
 \end{tabular}
\end{table}

Gas-phase calculations of radical-neutral reactions thus appear to be accurate enough to represent the very same reactions on an ice surface. This even holds for ices composed of water molecules as typical changes of the activation energy are roughly only 1-2~kJ/mol. 

Even though carbon monoxide constitutes a large molecular fraction, the expected weak interaction between CO and any of the species relevant here should not affect the reaction energetics considerably, as has also been seen for the reaction route \ce{H + CO} \citep{Rimola:2014}.

For barrierless reactions, on the other hand, a possible surface effect on the reaction progress is related to the orientation of molecules at the ice surface. Although the hydrogen bond of a S-H functional group is weaker than for O-H \citep{Biswal:2009}, there may still be an effect on the preferred binding mode of an \ce{H_nCS} molecule. Sampling a variety of binding sites and subsequently determining branching ratios per binding site is, however, beyond the scope of this work. Nonetheless, the branching ratios ($BR$) calculated here are the first to be reported in the literature and serve to show the major and minor product channels. They should therefore be regarded as a zero-order approximation to the `true' values on the surface. 

\section{Results and Discussion}
 
\subsection{Reaction network}
\begin{figure}[t]
  \includegraphics[width=9cm]{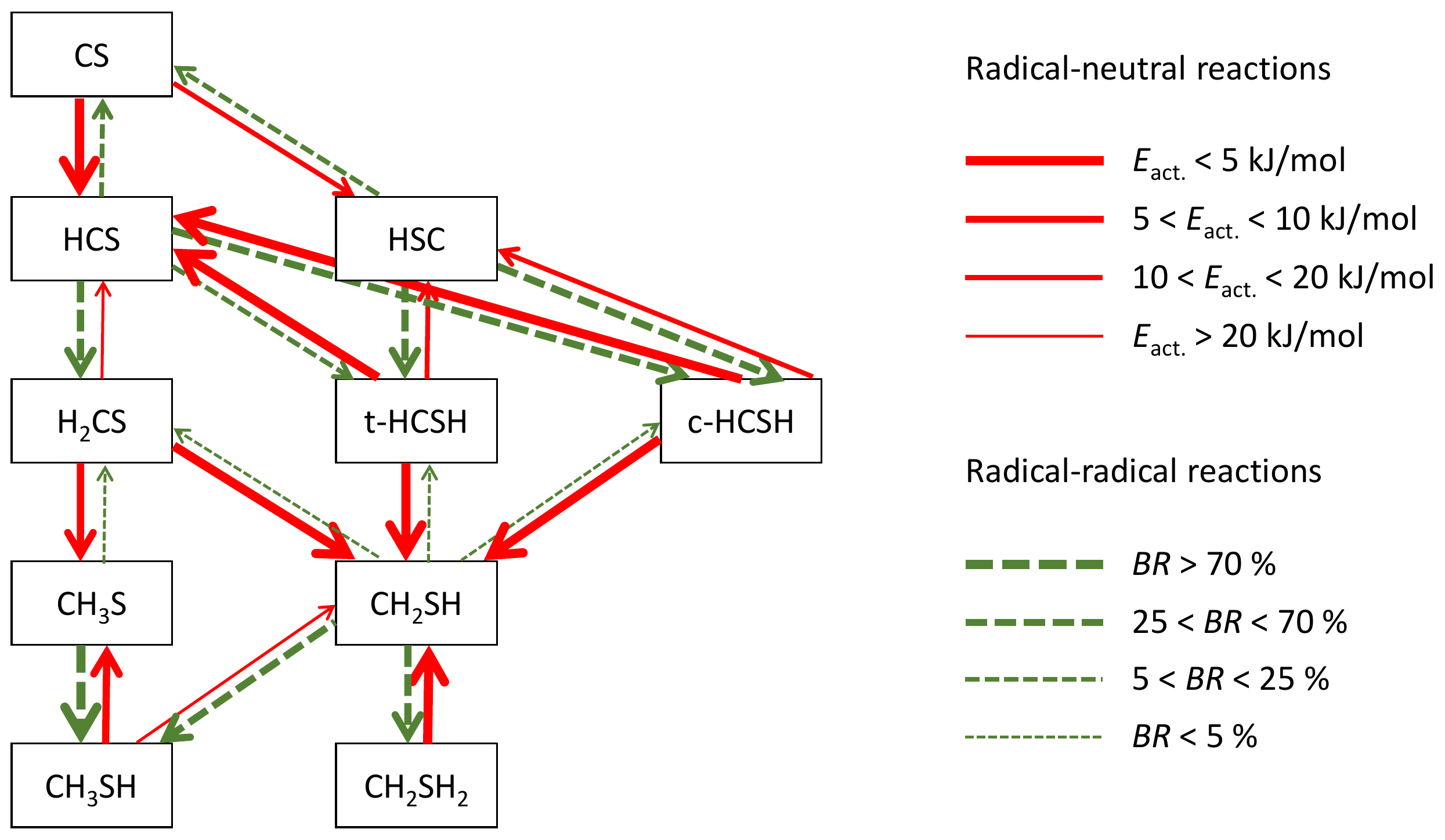}
  \caption{Reaction network connecting CS and \ce{CH3SH} via various intermediate \ce{H_nCS} species with an indication for the relevant activation energies and branching ratios. Red solid lines indicate reactions between a radical (H) and a neutral \ce{H_nCS} species (n = 0, 2, 4), for which transition states can be found connecting the two minima. Green, dashed lines, on the other hand, indicate barrierless radical-radical reactions. The thickness of the arrows corresponds to the likelihood of a particular reaction to take place.}
  \label{network}
\end{figure}

In Fig.~\ref{network}, the reaction network for hydrogen addition and abstraction reactions is depicted. The species shown on the left-hand side are the most stable and the relative energy increases towards the right-hand side. Isomerization reactions connecting species in the same row have been omitted for clarity. All activation energies that we found for these reactions exceed values of {100} kJ/mol (see also \citet{Kobayashi:2011}). Unless actively catalysed by surface molecules \citep{Tachikawa:2016}, these can be currently regarded as highly unlikely to occur at low temperatures in interstellar ices.  

\begin{table*}[t]
 \centering
 \caption{{Vibrationally adiabatic activation energy ($E_\text{act.}$) and reaction energy ($E_\text{react.}$) with respect to the separated reactants, calculated crossover temperature ($T_\text{c}$), the calculated instanton rate constant extrapolated to low temperature ($k_\text{low-T}$) and rectangular barrier rate constant ($k_\text{rect. barr.}$) for the radical-neutral reactions. The activation energies in parentheses are calculated with respect to the PRC and thus correspond to the unimolecular rate constants.}}\label{allbarrier}
 \begin{tabular}{lr@{ }r  lrrll}
  Reaction                              & \multicolumn{2}{r}{$E_\text{act.}$ $^{a}$} & {$E_\text{act.}$ (KIDA) $^{b}$} & $T_\text{c}$ $^{a}$ & $E_\text{react.} $ $^{a}$       & $k_\text{low-T}$ $^{a}$       & {$k_\text{rect. barr.}$ $^{c}$}  \\
                                        & \multicolumn{2}{r}{(kJ/mol)}  & {(kJ/mol)} & (K)         & (kJ/mol)              & (s$^{-1}$)             & (s$^{-1}$) \\
  \ce{H + CS -> HCS}                    & 3.6   & (3.1)         & 8.3  &  79             & -224.2                & $1.1 \times 10^{10}$  & $3.8 \times 10^{8}$ \\
  \ce{H + CS -> HSC}                    & 13.2  & (12.1)        & --  &  150             & -63.2                 & $4.3 \times 10^6$     & $1.8 \times 10^{5}$ \\
  \ce{H + H2CS -> H2 + HCS}             & 28.8  & (28.1)        & --  &  384             & -46.0                 & $1.3 \times 10^4$     & $5.1 \times 10^{1}$ \\
  \ce{H + H2CS -> CH3S}                 & 7.5   & (5.8)         & 10.0  &  103            & -222.0                & $1.3 \times 10^9$     & $2.1 \times 10^{7}$ \\
  \ce{H + H2CS -> CH2SH}                & 3.1   & (1.6)         & 6.7  &  67             & -190.6                & $1.5 \times 10^{11}$  & $3.5 \times 10^{9}$ \\
  \ce{H + trans\;HCSH -> H2 + HCS}      & 3.8   & (3.1)         & --  &  131             & -227.6                & $2.2 \times 10^{10}$  & $3.8 \times 10^{8}$ \\
  \ce{H + trans\;HCSH -> H2 + HSC}      & 16.9  & (15.6)        & --  &  330             & -66.5                 & $6.4 \times 10^6$     & $2.1 \times 10^{4}$ \\
  \ce{H + trans\;HCSH -> CH2SH }        & 0.0   & (0.0)         & --  &  --              & -372.1                & barrierless           & $1.0 \times 10^{12}$ \\
  \ce{H + cis\;HCSH -> H2 + HCS}        & 1.7   & (0.7)         & --  &  67              & -232.8                & $4.9 \times 10^{11}$  & $2.4 \times 10^{10}$ \\
  \ce{H + cis\;HCSH -> H2 + HSC}        & 12.5  & (11.3)        & --  &  270             & -71.8                 & $< 2.1 \times 10^7$   & $3.0 \times 10^{5}$ \\
  \ce{H + cis\;HCSH -> CH2SH }          & 0.0   & (0.0)         & --  &  --              & -377.4                & barrierless           & $1.0 \times 10^{12}$ \\
  \ce{H + CH3SH -> H2 + CH3S}           & 9.5   & (8.7)         & 6.7  &  206            & -68.9                 & $1.0 \times 10^8$     & $1.9 \times 10^{6}$ \\
  \ce{H + CH3SH -> H2 + CH2SH}          & 34.0  & (33.4)        & --  &  353             & -37.5                 & $1.9 \times 10^2$      & $6.0 \times 10^{0}$ \\
  \ce{H + CH3SH -> H2S + CH3}           & 16.8  & (16.2)        & 13.3  &  127            & -63.4                 & $< 4.2 \times 10^2$   & $1.5 \times 10^{4}$         \\
  \ce{H + CH2SH2 -> H2 + CH2SH}         & 0.0   & (0.0)         & --  &  --              & -293.5                & barrierless           & $1.0 \times 10^{12}$ \\
  \multicolumn{8}{l}{   $^{a}$ calculated with MPWB1K/def2-TZVP} \\
  \multicolumn{8}{l}{  {$^{b}$ as obtained from the KIDA database} } \\
  \multicolumn{8}{l}{  {$^{c}$ calculated using the unimolecular $E_\text{act.}$, a barrier width of 1 $\AA$, and a trial frequency of 10$^{12}$ s$^{-1}$} } \\  
  \end{tabular}
\end{table*}

\begin{table*}[t]
 \centering
 \caption{Vibrationally adiabatic reaction energy ($E_\text{react.}$) with respect to the separated reactants and corresponding branching ratio ($BR$) for the radical-radical reactions. }\label{allbarrierless}
 \begin{tabular}{lrl  l}
  Reaction                      & $E_\text{react.}$  $^{a}$     & $BR$  $^{b}$ & $BR$ (KIDA) $^{c}$\\  
                                & (kJ/mol)                      & (\%) & (\%) \\
  \ce{H + HCS -> H2 + CS}       & -196.2                        & 20            & -- \\
  \ce{H + HCS -> H2CS}          & -374.4                        & 35            & 100 \\
  \ce{H + HCS -> trans\;HCSH}   & -192.8                        & 15            & -- \\
  \ce{H + HCS -> cis\;HCSH}     & -187.6                        & 30            & -- \\
  \ce{H + HSC -> H2 + CS}       & -357.2                        & 20 $^{d}$     & -- \\
  \ce{H + HSC -> trans\;HCSH}   & -353.8                        & 30 $^{d}$     & -- \\
  \ce{H + HSC -> cis\;HCSH}     & -348.6                        & 30 $^{d}$     & -- \\
  \ce{H + CH3S -> H2 + H2CS}    & -198.4                        & 0 $^{e}$      & -- \\
  \ce{H + CH3S -> CH3SH}        & -351.5                        & 75 $^{e}$     & 100 \\
  \ce{H + CH2SH -> H2 + H2CS}           & -229.8                & 0-5   $^{f}$  & -- \\
  \ce{H + CH2SH -> H2 + trans\;HCSH}    & -48.3                 & 0     $^{f}$  & -- \\
  \ce{H + CH2SH -> H2 + cis\;HCSH}      & -43.0                 & 0     $^{f}$  & -- \\
  \ce{H + CH2SH -> CH2SH2}              & -126.8                & 30-35 $^{f}$  & -- \\  
  \ce{H + CH2SH -> CH3SH}               & -382.9                & 45-55 $^{f}$         & 100 \\
  \multicolumn{4}{l}{   $^{a}$ calculated with MPWB1K/def2-TZVP} \\
  \multicolumn{4}{l}{   $^{b}$ calculated both with B3LYP/def2-TZVP and PBEh-3c/mSVP. } \\
  \multicolumn{4}{l}{{$^{c}$ as obtained from the KIDA database}} \\
  \multicolumn{4}{l}{   $^{d}$ 20\% did not lead to a reaction} \\
  \multicolumn{4}{l}{   $^{e}$ 25\% did not lead to a reaction} \\
  \multicolumn{4}{l}{   $^{f}$ 5-25\% did not lead to a reaction} \\
  \end{tabular}
\end{table*}
All corresponding activation and reaction energies are reported in Tables~\ref{allbarrier} and~\ref{allbarrierless}. These have been calculated with respect to the separated reactants, although for the activation energies, values with respect to the PRC are also reported. The latter correspond directly to the calculated rate constants. {An overview of the energy profile diagram is depicted in Fig.~\ref{PES} in Appendix~\ref{Profile}.}

For the reactions that occur via a barrier, unimolecular instanton rate constants have been calculated at temperatures lower than the respective crossover temperatures, $T_\text{c}$, down to {45}~K with canonical instanton theory. The crossover temperature is defined as $ {\hbar \omega_b}/{2\pi k_\text{B}} $ , where $\omega_b$ is the absolute value of the imaginary frequency at the transition state, $\hbar$ Planck's constant divided by $2\pi$, and $k_\text{B}$ Boltzmann's constant; it indicates the temperature below which tunneling dominates the reaction mechanism. 
Furthermore, in Table~\ref{allbarrier}, the recommended low-temperature values for the rate constant are reported. This choice is explained in Appendix~\ref{RC} along with the temperature-dependent values for the rate constant. 
For both addition and abstraction reactions, within each category, the height of the barrier is directly related to the rate constant. 
We note that very low barriers result in $k_\text{low-T}$ approaching $10^{12}$ s$^{-1}$, while high barriers lead to values around $10^2$ s$^{-1}$. In general, many of the rate constants are high ($> 10^6$ s$^{-1}$), which means that they are faster than or competitive with hydrogen atom diffusion;  \cite{Senevirathne:2017,Asgeirsson:2017}. Finally, breaking the C-S bond at low temperature via the reaction \ce{H + CH3SH -> H2S + CH3} is six orders of magnitude slower than \ce{CH3S} formation as a result of the much lower efficiency of heavy-atom tunneling. This holds even though the activation energy is similar to \ce{H + trans\;HCSH -> H2 + HSC}.

\begin{figure*}
 \centering
 \includegraphics[width=3.9cm]{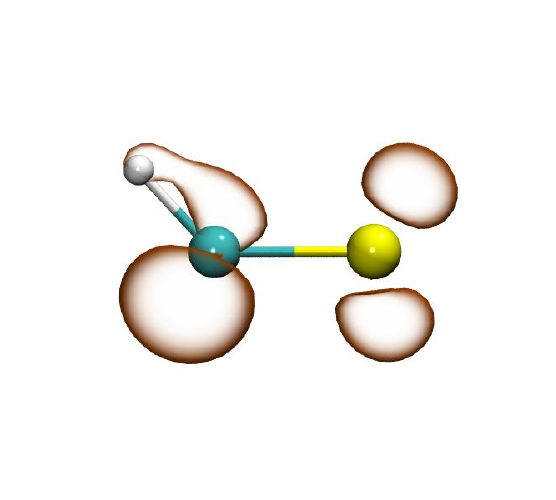} 
 \includegraphics[width=3.9cm]{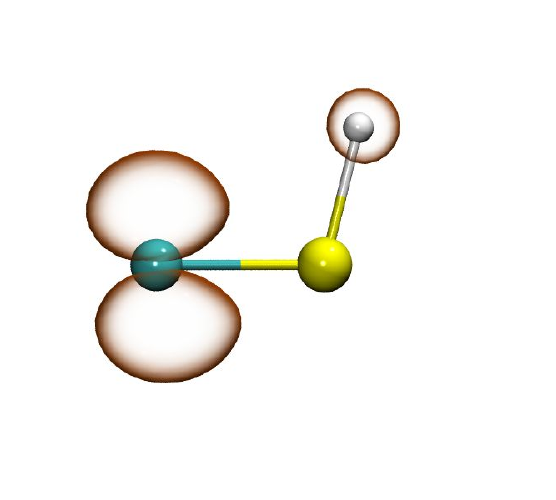}
 \includegraphics[width=3.9cm]{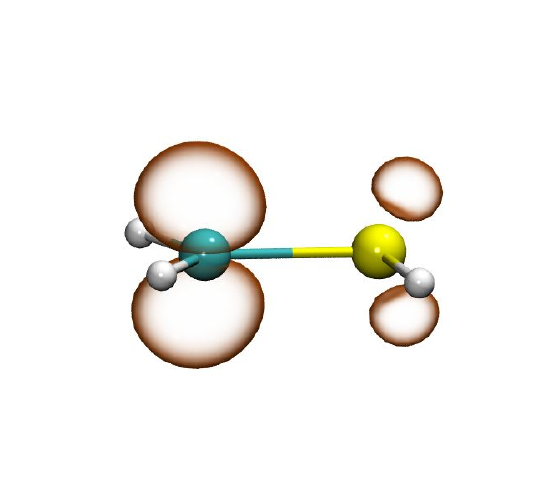}
 \includegraphics[width=3.9cm]{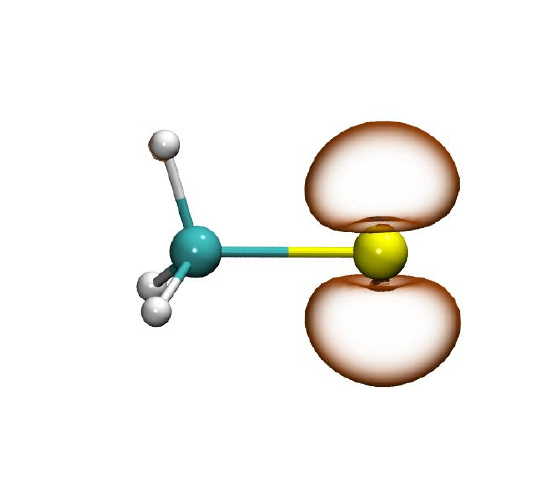} 
 \caption{Spin density of the \ce{H_nCS} radicals involved in the reaction network (isoval = 0.015).}\label{spindens}
\end{figure*}

For the radical-radical reactions, branching ratios have been determined in the gas phase. {The general trends are summarised in Table~\ref{allbarrierless}. We note that only for \ce{H + CH2SH} did results differ between the two functionals.} Although the distance between the H atom and the C-S bond is relatively small, a number of optimisation attempts resulted in the formation of a dimer structure, where no reaction had taken place. This can be clarified with the help of spin densities, that is, a surplus of alpha over beta spin orbitals, calculated for all four radicals and depicted in Fig.~\ref{spindens}. For \ce{HCS}, there is spin density present around all atoms and therefore four product channels are available. For the radical \ce{CH3S}, the opposite situation can be found, where the majority of spin density is located on the sulphur atom and consequently most optimisations result in the formation of \ce{CH3SH}. For runs where the H-atom is placed close to the methyl group, a non-reactive dimer is formed.

\subsection{Astrochemical implications}
Combining the insight obtained from both the rate constant and branching ratio calculations, it becomes clear that some species will be present at higher concentrations and for a longer time in ices. Specifically, the radical \ce{HCS} is not only primarily formed through the reaction \ce{H + CS}, but is also the main product of H-abstraction reactions from both trans- and cis-\ce{HCSH}. As a result of this prolonged presence in the ice, the radical is also available for other reactions, possibly being incorporated into larger species via carbon-carbon bonds. Going one step further in the reaction network, H additions to \ce{H2CS} and trans- and cis-\ce{HCSH} are likely to result in a significant amount of \ce{CH2SH}. The reaction \ce{H + H2CS -> CH2SH} is more likely to take place than the competing \ce{H + H2CS -> CH3S}. The radical \ce{CH3S} on the other hand is the main product of the reaction between H and \ce{CH3SH}. Finally, once either \ce{CH3S} or \ce{CH2SH} has been formed, it is quite unlikely for H-abstraction reactions to be able to reduce the number of hydrogen atoms attached to the C-S centre to less than three.  

Within the framework of reaction kinetics and the implementation thereof in astrochemical models, first it should be pointed out that the values for the rate constants presented here are valid in the low-temperature regime.  
{The KIDA database \citep{Wakelam:2012} and in particular \citet{Vidal:2017} provide barrier heights (activation energies) and branching ratios, also listed in Tables~\ref{allbarrier} and~\ref{allbarrierless}. These values were estimated from calculations performed at the M06-2X/cc-pVTZ level of theory with a few additional calculations at the MP2/cc-pVTZ level of theory. Here, we make use of MPWB1K/def2-TZVP, which was explicitly benchmarked against two high-level theories, namely both CCSD(T)-F12/cc-VTZ-F12 and MRCI(+Q)-F12/VTZ-F12 single-point energies. Moreover, the values provided here are in good agreement with previous work by \citet{Kobayashi:2011, Kerr:2015}. Furthermore, the branching ratios given by KIDA were determined solely by the most exothermic reaction channel. Here, on the other hand, the orientation of the incoming hydrogen atom with respect to the CS backbone is also taken into account. Therefore, the values presented here are thought to be of a higher accuracy than those currently implemented in models. In particular, the first reactions of the reaction network, \ce{H + CS} and \ce{H + H2CS}, will proceed much faster than currently taken into account by, for example, \citet{Vidal:2017}, whereas abstraction from \ce{CH3SH} will be slower. This may lead to a higher amount of methanethiol being built up. }

In Table~\ref{allbarrier}, two types of calculated rate constants are also listed: using instanton theory and the rectangular barrier approximation. The latter is commonly used in astrochemical models, based on early approaches by \citet{Tielens:1982} and \citet{Hasegawa:1993}. Obviously the rectangular barrier approximation is a significant improvement on the use of classically calculated rate constants with (harmonic) transition state theory. However, the order of magnitude may still differ with respect to values calculated by more accurate methods, here resulting in values that are lower by up to a factor of $\sim$250. This does not only affect the rate of the reactions in the models, and therefore the build-up of species in the ice mantles, but it also relates to the competition of reactions with diffusion of hydrogen atoms. Rate constants for hydrogen diffusion have been shown to span a large range of values, with the exact value depending on the specific binding site \cite{Senevirathne:2017,Asgeirsson:2017}. {Therefore, it is advisable to directly use the values for the rate constants calculated by means of instanton theory at low temperature, instead of rate constants calculated with the rectangular barrier approximation, regardless of the quality of the activation energy used within that expression. }

{Although the upcoming James Webb Space Telescope provides an increased peak-to-noise ratio, observations of CS-containing molecules in the solid phase may only be fully possible if extensive spectral data become available for different mixing ratios. These data will be difficult to obtain experimentally as a result of the hazardous characteristics of the molecules investigated here. The use of calculated spectra, based, for example, on density functional theory, may in fact be of great assistance. }

\section{Conclusion}
In summary, a total of 29 reactions have been calculated within the reaction network starting from the hydrogenation of carbon monosulfide leading to the formation of methyl mercaptan. Extending on previous work, we have found that not only can \ce{HCS}, \ce{H2CS}, \ce{CH2SH}, \ce{CH3S}, and \ce{CH3SH}  be formed, but additionally \ce{HSC}, {t-HCSH}, and {c-HCSH} appear naturally, as well as \ce{H2S}, \ce{CH3}, and \ce{CH2SH2} to a lesser extent. For all 12 reactions with a barrier, rate constants have been calculated and the values at 45~K can be used in astrochemical models as an extrapolation for even lower temperatures. For the radical-radical reactions, preliminary branching ratios are provided as well, indicating the most likely product channels. In fact, not all reactions that one would a priori imagine to take place are actually possible due to the distribution of spin density.

 Early-on in  the reaction network, \ce{HCS} is expected to be the main radical, whereas \ce{CH2SH} dominates at later stages. Eventually, H-abstraction from \ce{CH3SH} can also take place, which results preferentially in \ce{CH3S}. Furthermore, it is important to note though that not only \ce{H2CS} but also trans- and cis-\ce{HCSH} can easily be formed. Finally, once three hydrogen atoms have been added to the original CS molecule, it is unlikely for H-abstraction to regenerate \ce{H_nCS} species with n $\leq$\ 2.      

The data presented here need to be included in a full gas-grain astrochemical model in order to appreciate the subtle differences that may occur through the competition with diffusion and/or other reactions.
The effect of surface molecules on the activation energies of the radical-neutral reactions studied is argued to be expected to be well within the chemical accuracy. However, the orientation of the molecules and radicals on the surface may inhibit and/or promote specific reactions to take place. How exothermicity, amorphicity, and ice composition influence the binding modes and possible reactions or branching ratios is not known. Future (theoretical) studies on these topics are paramount.

\begin{acknowledgements}
Yuri Alexandre Aoto and Johannes K\"astner are gratefully acknowledged for useful discussions. Gleb Fedoseev, Jan Meisner, and J\"{o}rg Meyer are thanked for proofreading.
The author acknowledges support for computer time by the state of Baden-W\"{u}rttemberg through bwHPC and the Germany Research Foundation (DFG) through grant no. INST 40/467-1FUGG. 
This project was financially supported by the European Union's Horizon 2020 research and innovation programme (grant agreement No. 646717, TUNNELCHEM), the Alexander von Humboldt Foundation and the Netherlands Organisation for Scientific Research (NWO) via a VENI fellowship (722.017.008)
\end{acknowledgements}


\bibliographystyle{aa}
\bibliography{all.bib}

\begin{thebibliography}{68}
\expandafter\ifx\csname natexlab\endcsname\relax\def\natexlab#1{#1}\fi

\bibitem[{Adler {et~al.}(2007)Adler, Knizia, \& Werner}]{Adler:2007}
Adler, T.~B., Knizia, G., \& Werner, H.-J. 2007, J. Chem. Phys., 127, 221106

\bibitem[{Ahlrichs {et~al.}(1989)Ahlrichs, B\"{a}r, H\"{a}ser, Horn, \&
  K\"{o}lmel}]{Ahlrichs:1989}
Ahlrichs, R., B\"{a}r, M., H\"{a}ser, M., Horn, H., \& K\"{o}lmel, C. 1989,
  Chem.~Phys.~Lett., 162, 165

\bibitem[{An {et~al.}(2017)An, Sellgren, Boogert, Ram\'{i}rez, \&
  Pyo}]{Deokkeun:2017}
An, D., Sellgren, K., Boogert, A. C.~A., Ram\'{i}rez, S.~V., \& Pyo, T.-S.
  2017, Astrophys.~J.~Lett., 843, L36

\bibitem[{\'Asgeirsson {et~al.}(2017)\'Asgeirsson, J\'onsson, \&
  Wikfeldt}]{Asgeirsson:2017}
\'Asgeirsson, V., J\'onsson, H., \& Wikfeldt, K.~T. 2017, J.~Phys.~Chem.~C,
  121, 1648

\bibitem[{Becke(1993)}]{Becke:1993}
Becke, A.~D. 1993, \jcp, 98, 5648

\bibitem[{Biswal {et~al.}(2009)Biswal, Shirhatti, \& Wategaonkar}]{Biswal:2009}
Biswal, H.~S., Shirhatti, P.~R., \& Wategaonkar, S. 2009, \jcpa, 113, 5633

\bibitem[{Boogert {et~al.}(2015)Boogert, Gerakines, \& Whittet}]{Boogert:2015}
Boogert, A.~A., Gerakines, P.~A., \& Whittet, D.~C. 2015,
  Ann.~Rev.~Astron.~Astrophys., 53, 541

\bibitem[{Callan \& Coleman(1977)}]{Callan:1977}
Callan, C.~G. \& Coleman, S. 1977, Phys.~Rev.~D, 16, 1762

\bibitem[{{Calmonte} {et~al.}(2016){Calmonte}, {Altwegg}, {Balsiger},
  {Berthelier}, {Bieler}, {Cessateur}, {Dhooghe}, {van Dishoeck}, {Fiethe},
  {Fuselier}, {Gasc}, {Gombosi}, {H{\"a}ssig}, {Le Roy}, {Rubin}, {S{\'e}mon},
  {Tzou}, \& {Wampfler}}]{Calmonte:2016}
{Calmonte}, U., {Altwegg}, K., {Balsiger}, H., {et~al.} 2016, \mnras, 462, S253

\bibitem[{Coleman(1977)}]{Coleman:1977}
Coleman, S. 1977, Phys.~Rev.~D, 15, 2929

\bibitem[{Deegan \& Knowles(1994)}]{Deegan:1994}
Deegan, M. J.~O. \& Knowles, P.~J. 1994, Chem. Phys. Lett., 227, 321

\bibitem[{Drozdovskaya {et~al.}(2018)Drozdovskaya, van Dishoeck, J\o{}rgensen,
  Calmonte, van~der Wiel, Coutens, Calcutt, M\"{u}ller, Bjerkeli, Persson,
  Wampfler, \& Altwegg}]{Drozdovskaya:2018}
Drozdovskaya, M.~N., van Dishoeck, E.~F., J\o{}rgensen, J.~K., {et~al.} 2018,
  \mnras, accepted

\bibitem[{Du \& Zhang(2017)}]{Du:2017}
Du, B. \& Zhang, W. 2017, Comp.~Theor.~Chem., 1117, 235

\bibitem[{Fedoseev {et~al.}(2015)Fedoseev, Cuppen, Ioppolo, Lamberts, \&
  Linnartz}]{Fedoseev:2015}
Fedoseev, G., Cuppen, H.~M., Ioppolo, S., Lamberts, T., \& Linnartz, H. 2015,
  \mnras, 448, 1288

\bibitem[{{Fuchs} {et~al.}(2009){Fuchs}, {Cuppen}, {Ioppolo}, {Romanzin},
  {Bisschop}, {Andersson}, {van Dishoeck}, \& {Linnartz}}]{Fuchs:2009}
{Fuchs}, G.~W., {Cuppen}, H.~M., {Ioppolo}, S., {et~al.} 2009, \aap, 505, 629

\bibitem[{{Garrod} \& {Herbst}(2006)}]{Garrod:2006a}
{Garrod}, R.~T. \& {Herbst}, E. 2006, Astron.~Astrophys., 457, 927

\bibitem[{Grimme {et~al.}(2015)Grimme, Brandenburg, Bannwarth, \&
  Hansen}]{Grimme:2015}
Grimme, S., Brandenburg, J.~G., Bannwarth, C., \& Hansen, A. 2015,
  J.~Chem.~Phys., 143, 054107

\bibitem[{Hasegawa \& Herbst(1993)}]{Hasegawa:1993}
Hasegawa, T.~I. \& Herbst, E. 1993, Mon.~Not.~R.~Astron.~Soc., 263, 589

\bibitem[{Hiraoka {et~al.}(1998)Hiraoka, Miyagoshi, Takayama, Yamamoto, \&
  Kihara}]{Hiraoka:1998}
Hiraoka, K., Miyagoshi, T., Takayama, T., Yamamoto, K., \& Kihara, Y. 1998,
  Astrophys.~J.~, 498, 710

\bibitem[{K\"astner(2014)}]{Kaestner:2014}
K\"astner, J. 2014, WIREs Comput. Mol. Sci., 4, 158

\bibitem[{K\"{a}stner {et~al.}(2009)K\"{a}stner, Carr, Keal, Thiel, Wander, \&
  Sherwood}]{Kaestner:2009}
K\"{a}stner, J., Carr, J.~M., Keal, T.~W., {et~al.} 2009, \jpca, 113, 11856

\bibitem[{Kerr {et~al.}(2015)Kerr, Alecu, Thompson, Gao, \&
  Marshall}]{Kerr:2015}
Kerr, K.~E., Alecu, I.~M., Thompson, K.~M., Gao, Y., \& Marshall, P. 2015,
  \jpca, 119, 7352, pMID: 25872011

\bibitem[{Knizia {et~al.}(2009)Knizia, Adler, \& Werner}]{Knizia:2009}
Knizia, G., Adler, T.~B., \& Werner, H.-J. 2009, J. Chem. Phys., 130, 054104

\bibitem[{Knowles {et~al.}(1993)Knowles, Hampel, \& Werner}]{Knowles:1993}
Knowles, P.~J., Hampel, C., \& Werner, H.-J. 1993, J. Chem. Phys., 99, 5219

\bibitem[{Knowles {et~al.}(2000)Knowles, Hampel, \& Werner}]{Knowles:err}
Knowles, P.~J., Hampel, C., \& Werner, H.-J. 2000, \jcp, 112, 3106

\bibitem[{Knowles \& Werner(1988)}]{Knowles:1988}
Knowles, P.~J. \& Werner, H.-J. 1988, Chem.~Phys.~Lett., 145, 514

\bibitem[{Kobayashi {et~al.}(2011)Kobayashi, Seki, Tanaka, \&
  Takayanagi}]{Kobayashi:2011}
Kobayashi, T., Seki, K., Tanaka, T., \& Takayanagi, T. 2011,
  Comp.~Theor.~Chem., 963, 256

\bibitem[{Kosztol\'{a}nyi {et~al.}(2003)Kosztol\'{a}nyi, Bak\'{o}, \&
  P\'{a}link\'{a}s}]{Kosztolanyi:2003}
Kosztol\'{a}nyi, T., Bak\'{o}, I., \& P\'{a}link\'{a}s, G. 2003, \jcp, 118,
  4546

\bibitem[{{Lamberts} \& {K{\"a}stner}(2017)}]{Lamberts:2017}
{Lamberts}, T. \& {K{\"a}stner}, J. 2017, \apj, 846, 43

\bibitem[{Lamberts \& K\"{a}stner(2017)}]{Lamberts:2017B}
Lamberts, T. \& K\"{a}stner, J. 2017, acc. \jpca, DOI 10.1021/acs.jpca.7b10296

\bibitem[{{Lamberts} {et~al.}(2016){Lamberts}, {Samanta}, {K{\"o}hn}, \&
  {K{\"a}stner}}]{Lamberts:2016B}
{Lamberts}, T., {Samanta}, P.~K., {K{\"o}hn}, A., \& {K{\"a}stner}, J. 2016,
  \pccp, 18, 33021

\bibitem[{Langer(1967)}]{Langer:1967}
Langer, J. 1967, Ann.~Phys., 41, 108

\bibitem[{Lee {et~al.}(1988)Lee, Yang, \& Parr}]{Lee:1988}
Lee, C., Yang, W., \& Parr, R.~G. 1988, Phys.~Rev.~B, 37, 785

\bibitem[{{Linke} {et~al.}(1979){Linke}, {Frerking}, \&
  {Thaddeus}}]{Linke:1979}
{Linke}, R.~A., {Frerking}, M.~A., \& {Thaddeus}, P. 1979, \apjl, 234, L139

\bibitem[{{Liszt} {et~al.}(1974){Liszt}, {Wilson}, {Penzias}, {Jefferts},
  {Wannier}, \& {Solomon}}]{Liszt:1974}
{Liszt}, H.~S., {Wilson}, R.~W., {Penzias}, A.~A., {et~al.} 1974, \apj, 190,
  557

\bibitem[{Majumdar {et~al.}(2016)Majumdar, Gratier, Vidal, Wakelam, Loison,
  Hickson, \& Caux}]{Majumdar:2016}
Majumdar, L., Gratier, P., Vidal, T., {et~al.} 2016, \mnras, 458, 1859

\bibitem[{{Marcelino} {et~al.}(2005){Marcelino}, {Cernicharo}, {Roueff},
  {Gerin}, \& {Mauersberger}}]{Marcelino:2005}
{Marcelino}, N., {Cernicharo}, J., {Roueff}, E., {Gerin}, M., \&
  {Mauersberger}, R. 2005, \apj, 620, 308

\bibitem[{Meisner {et~al.}(2017)Meisner, Lamberts, \& K\"astner}]{Meisner:2017}
Meisner, J., Lamberts, T., \& K\"astner, J. 2017, ACS Earth Space Chem.

\bibitem[{Metz {et~al.}(2014)Metz, K\"astner, Sokol, Keal, \&
  Sherwood}]{Metz:2014}
Metz, S., K\"astner, J., Sokol, A.~A., Keal, T.~W., \& Sherwood, P. 2014, WIREs
  Comput. Mol. Sci., 4, 101

\bibitem[{Miller(1975)}]{Miller:1975}
Miller, W.~H. 1975, J.~Chem.~Phys., 62, 1899

\bibitem[{M\"uller {et~al.}(2016)M\"uller, {Belloche, Arnaud}, {Xu, Li-Hong},
  {Lees, Ronald M.}, {Garrod, Robin T.}, {Walters, Adam}, {van Wijngaarden,
  Jennifer}, {Lewen, Frank}, {Schlemmer, Stephan}, \& {Menten, Karl
  M.}}]{Mueller:2016}
M\"uller, H. S.~P., {Belloche, Arnaud}, {Xu, Li-Hong}, {et~al.} 2016, \aap,
  587, A92

\bibitem[{Palumbo {et~al.}(1997)Palumbo, Geballe, \& Tielens}]{Palumbo:1997}
Palumbo, M.~E., Geballe, T.~R., \& Tielens, A. G. G.~M. 1997, \apj, 479, 839

\bibitem[{{Penzias} {et~al.}(1971){Penzias}, {Solomon}, {Wilson}, \&
  {Jefferts}}]{Penzias:1971}
{Penzias}, A.~A., {Solomon}, P.~M., {Wilson}, R.~W., \& {Jefferts}, K.~B. 1971,
  \apjl, 168, L53

\bibitem[{Peterson {et~al.}(2008)Peterson, Adler, \& Werner}]{Peterson:2008}
Peterson, K.~A., Adler, T.~B., \& Werner, H.-J. 2008, \jcp, 128, 084102

\bibitem[{Richardson(2016)}]{Richardson:2016}
Richardson, J.~O. 2016, J. Chem. Phys., 144, 114106

\bibitem[{Rimola {et~al.}(2014)Rimola, {Taquet, Vianney}, {Ugliengo, Piero},
  {Balucani, Nadia}, \& {Ceccarelli, Cecilia}}]{Rimola:2014}
Rimola, A., {Taquet, Vianney}, {Ugliengo, Piero}, {Balucani, Nadia}, \&
  {Ceccarelli, Cecilia}. 2014, \aap, 572, A70

\bibitem[{Rommel {et~al.}(2011)Rommel, Goumans, \& K\"{a}stner}]{Rommel:2011}
Rommel, J.~B., Goumans, T.~P.~M., \& K\"{a}stner, J. 2011, J. Chem. Theory
  Comput., 7, 690

\bibitem[{Rommel \& K\"{a}stner(2011)}]{Rommel:2011b}
Rommel, J.~B. \& K\"{a}stner, J. 2011, \jcp, 134, 184107

\bibitem[{Senevirathne {et~al.}(2017)Senevirathne, Andersson, Dulieu, \&
  Nyman}]{Senevirathne:2017}
Senevirathne, B., Andersson, S., Dulieu, F., \& Nyman, G. 2017,
  Mol.~Astrophys., 6, 59

\bibitem[{Sherwood {et~al.}(2003)Sherwood, de~Vries, Guest, Schreckenbach,
  Catlow, French, Sokol, Bromley, Thiel, Turner, Billeter, Terstegen, Thiel,
  Kendrick, Rogers, Casci, Watson, King, Karlsen, Sj{\o}voll, Fahmi,
  Sch{\"a}fer, \& Lennartz}]{Sherwood:2003}
Sherwood, P., de~Vries, A.~H., Guest, M.~F., {et~al.} 2003, J. Mol. Struct.
  (THEOCHEM), 632, 1

\bibitem[{Shiozaki {et~al.}(2011)Shiozaki, Knizia, \& Werner}]{Shiozaki:2011}
Shiozaki, T., Knizia, G., \& Werner, H.-J. 2011, J.~Chem.~Phys., 134, 034113

\bibitem[{Shiozaki \& Werner(2011)}]{Shiozaki:2011a}
Shiozaki, T. \& Werner, H.-J. 2011, J.~Chem.~Phys., 134, 184104

\bibitem[{{Sinclair} {et~al.}(1973){Sinclair}, {Fourikis}, {Ribes}, {Robinson},
  {Brown}, \& {Godfrey}}]{Sinclair:1973}
{Sinclair}, M.~W., {Fourikis}, N., {Ribes}, J.~C., {et~al.} 1973,
  Aust.~J.~Phys., 26, 85

\bibitem[{Stephens {et~al.}(1994)Stephens, Devlin, Chabalowski, \&
  Frisch}]{Stephens:1994}
Stephens, P.~J., Devlin, F.~J., Chabalowski, C.~F., \& Frisch, M.~J. 1994,
  \jpc, 98, 11623

\bibitem[{Tachikawa \& Kawabata(2016)}]{Tachikawa:2016}
Tachikawa, H. \& Kawabata, H. 2016, \jpca, 120, 6596

\bibitem[{{Tielens} \& {Hagen}(1982)}]{Tielens:1982}
{Tielens}, A. G. G.~M. \& {Hagen}, W. 1982, \aap, 114, 245

\bibitem[{Treutler \& Ahlrichs(1995)}]{Treutler:1995}
Treutler, O. \& Ahlrichs, R. 1995, J.~Chem.~Phys., 102, 346

\bibitem[{v.~Arnim \& Ahlrichs(1998)}]{Arnim:1998}
v.~Arnim, M. \& Ahlrichs, R. 1998, J.~Comp.~Chem., 19, 1746

\bibitem[{Valiev {et~al.}(2010)Valiev, Bylaska, Govind, Kowalski, Straatsma,
  Dam, Wang, Nieplocha, Apra, Windus, \& de~Jong}]{Valiev:2010}
Valiev, M., Bylaska, E., Govind, N., {et~al.} 2010, Comp.~Phys.~Comm., 181,
  1477

\bibitem[{Vasyunin \& Herbst(2013)}]{Vasyunin:2013}
Vasyunin, A.~I. \& Herbst, E. 2013, \apj, 762, 86

\bibitem[{Vidal {et~al.}(2017)Vidal, Loison, Jaziri, Ruaud, Gratier, \&
  Wakelam}]{Vidal:2017}
Vidal, T. H.~G., Loison, J.-C., Jaziri, A.~Y., {et~al.} 2017, \mnras, 469, 435

\bibitem[{{Wakelam} {et~al.}(2012){Wakelam}, {Herbst}, {Loison}, {Smith},
  {Chandrasekaran}, {Pavone}, {Adams}, {Bacchus-Montabonel}, {Bergeat},
  {B{\'e}roff}, {Bierbaum}, {Chabot}, {Dalgarno}, {van Dishoeck}, {Faure},
  {Geppert}, {Gerlich}, {Galli}, {H{\'e}brard}, {Hersant}, {Hickson},
  {Honvault}, {Klippenstein}, {Le Picard}, {Nyman}, {Pernot}, {Schlemmer},
  {Selsis}, {Sims}, {Talbi}, {Tennyson}, {Troe}, {Wester}, \&
  {Wiesenfeld}}]{Wakelam:2012}
{Wakelam}, V., {Herbst}, E., {Loison}, J.-C., {et~al.} 2012, \apjs, 199, 21

\bibitem[{Walsh {et~al.}(2016)Walsh, Loomis, \"{O}berg, Kama, van~'t Hoff,
  Millar, Aikawa, Herbst, Weaver, \& Nomura}]{Walsh:2016}
Walsh, C., Loomis, R.~A., \"{O}berg, K.~I., {et~al.} 2016, \apjl, 823, L10

\bibitem[{{Watanabe} \& {Kouchi}(2002)}]{Watanabe:2002}
{Watanabe}, N. \& {Kouchi}, A. 2002, \apjl, 571, L173

\bibitem[{Weigend {et~al.}(1998)Weigend, H{\"a}ser, Patzelt, \&
  Ahlrichs}]{Weigend:1998}
Weigend, F., H{\"a}ser, M., Patzelt, H., \& Ahlrichs, R. 1998, \cpl, 294, 143

\bibitem[{Werner \& Knowles(1988)}]{Werner:1988}
Werner, H. \& Knowles, P.~J. 1988, J.~Chem.~Phys., 89, 5803

\bibitem[{Werner {et~al.}(2015)Werner, Knowles, Knizia, Manby, {Sch\"{u}tz},
  Celani, Gy\"orffy, Kats, Korona, Lindh, Mitrushenkov, Rauhut, Shamasundar,
  Adler, Amos, Bernhardsson, Berning, Cooper, Deegan, Dobbyn, Eckert, Goll,
  Hampel, Hesselmann, Hetzer, Hrenar, Jansen, K\"oppl, Liu, Lloyd, Mata, May,
  McNicholas, Meyer, Mura, Nicklass, O'Neill, Palmieri, Peng, Pfl\"uger,
  Pitzer, Reiher, Shiozaki, Stoll, Stone, Tarroni, Thorsteinsson, \&
  Wang}]{MOLPRO}
Werner, H.-J., Knowles, P.~J., Knizia, G., {et~al.} 2015, MOLPRO, version
  2015.1, a package of ab initio programs, see http://www.molpro.net

\bibitem[{Zhao \& Truhlar(2004)}]{Zhao:2004}
Zhao, Y. \& Truhlar, D.~G. 2004, \jcpa, 108, 6908

\end{thebibliography}

\appendix

\section{Computational Details}\label{CompDet}

\subsection{Radical-neutral reactions}

{For all radical-neutral reactions listed in Table~\ref{allbarrier}, full geometry optimisations for the pre-reactive complex, the reactants and products, and a transition state optimisation have been performed including the calculation of their respective Hessians, followed by a rate constant calculation. } The optimizations were verified by the appropriate number of imaginary frequencies, that is, zero for the reactants, products, and pre-reactive complexes, and one for the transition states. These, as well as all instanton rate constant calculations, have been performed using MPWB1K/def2-TZVP \citep{Zhao:2004,Weigend:1998}. The DL-find library \citep{Kaestner:2009} within the Chemshell framework \citep{Sherwood:2003, Metz:2014} was used in combination with NWChem version 6.6 \citep{Valiev:2010}. No additional dispersion correction was taken into account; see \citet{Zhao:2004}. Furthermore, performance has been compared to single-point energies at the CCSD(T)-F12a/cc-VTZ-F12 level \citep{Knowles:1993, Knowles:err, Deegan:1994,Adler:2007, Knizia:2009,Peterson:2008} and, where needed as indicated by D1 and T1 diagnostics, at the MRCI(+Q)-F12/cc-VTZ-F12 level of theory  \citep{Werner:1988,Knowles:1988,Shiozaki:2011,Shiozaki:2011a,Peterson:2008} with Molpro 2012 \citep{MOLPRO}; see Appendix~\ref{Bench}. In the latter case, a Davidson correction was applied using a relaxed reference. In short, because of the thorough benchmark, the activation energies are best seen to be chemically accurate (error $<$ 4 kJ/mol or 480~K). Pre-reactive complex and transition state geometries are listed in Appendix~\ref{Geom}.

Reaction rate constants were {subsequently} calculated using instanton theory \citep{Miller:1975,Langer:1967,Coleman:1977,Callan:1977,Rommel:2011,Kaestner:2014,Richardson:2016}.
{Instanton theory, or the imaginary-F method, makes use of statistical Feynman path integral theory to take quantum effects of atomic movements into account. Both the partition function of the reactant state as well as that of the instanton path, the transition-state equivalent, are obtained by a steepest-descent approximation to the phase space integrals. First, a discretized Feynman path is optimised with the Newton-Raphson method to find the instanton, which is a first-order saddle point in the space of closed Feynman paths \citep{Rommel:2011,Rommel:2011b}. Secondly, the Hessians of the potential energy at all images of the Feynman path are calculated to be able to evaluate the rate constant. Instanton theory is generally considered to be more accurate than one-dimensional tunneling corrections, like Eckart or Bell corrections. More information can be found in \citet{Kaestner:2014}.}
Instanton calculations are considered to be converged when all components of the nuclear gradient are smaller than $1 \times 10^{-8}$ a.u.. The instanton is discretized over 60 images and convergence with respect to the number of images has been checked for six reactions by optimising the low-temperature rate constant at 45~K with 118 images. The deviation concerned at most a factor 1.5.
Partition functions are calculated within the rigid-rotor-harmonic-oscillator approximation. The translational temperature is equal to the overall temperature of the reaction system as excess heat is removed instantaneously and the thermal equilibrium is assumed throughout the whole reaction. Restricted rotation on the surface is taken into account by keeping the rotational partition function constant for the reactant and transition state \citep{Meisner:2017, Lamberts:2017B}. Furthermore, unimolecular rate constants are calculated, representing the situation where two reactants diffuse on a surface and form a pre-reactive complex (PRC) prior to reaction. The rate constant essentially describes the decay of the PRC.

\subsection{Radical-radical reactions}
 
{For each of the radical-radical reactions listed in Table~\ref{allbarrierless}, calculations consisted of unrestricted symmetry-broken geometry optimisations for 60 different initial geometries along with spin density calculations,} performed using Turbomole version 7.1 \citep{Ahlrichs:1989,Treutler:1995,Arnim:1998}. Two relatively cheap functional and basis set combinations were used: both B3LYP/def2-TZVP \citep{Becke:1993,Lee:1988,Stephens:1994,Weigend:1998} and PBEh-3c/def2-mSVP \citep{Grimme:2015}, again using the DL-find library within the Chemshell framework. {The primary goal was to obtain the major and minor product channels of reactions between a H atom and an \ce{H_nCS} radical (n = 1, 3).} Branching ratios, $BR$, are determined by placing the \ce{H_nCS} radical at the origin and selecting 60 regularly distributed positions on a sphere of radius {3--4 {\AA}} with respect to the centre of the C-S bond. These positions serve as initial guesses for a geometry optimisation routine. The final optimised geometry then indicates which products have been formed, and the total number of occurrences of a particular final geometry divided by 60 is defined as $BR$. No Hessian calculations were performed. 
{Each geometry optimisation was performed twice, once for each functional, where the same input geometries were used for both calculations. The use of two different functionals serves as a simple double-check for the validity of the calculated branching ratio without
significantly increasing the computational cost. Appendix~\ref{Bench} provides the reaction energies calculated with both functionals compared to those calculated with MPWB1K/def2-TZVP and CCSD(T)-F12/cc-VTZ-F12. All levels of theory show that the reactions are exothermic and, moreover, that most reaction energies differ by less than $\sim$ 10\%. Therefore the assumption is made that this approach is valid for the qualitative study performed here. To obtain better quantification, at least CASSCF calculations would be necessary. For some selected initial geometries, we performed CASSCF/VDZ tests and indeed found the same products as for the DFT approach. }  Note that again  excess heat is removed instantaneously and the thermal equilibrium is assumed throughout the whole reaction. 

\section{Benchmark}\label{Bench}

From here onwards, the following reaction labelling is used for the radical-neutral reactions:
\begin{itemize}
\item R1: \ce{H + CS -> HCS}                    
\item R2: \ce{H + CS -> HSC}                    
\item R3: \ce{H + H2CS -> H2 + HCS}             
\item R4: \ce{H + H2CS -> CH3S}                 
\item R5: \ce{H + H2CS -> CH2SH}                
\item R6: \ce{H + trans\;HCSH -> H2 + HCS}      
\item R7: \ce{H + trans\;HCSH -> H2 + HSC} 
\item R8: \ce{H + trans\;HCSH -> CH2SH }
\item R9: \ce{H + cis\;HCSH -> H2 + HCS}        
\item R10: \ce{H + cis\;HCSH -> H2 + HSC} 
\item R11: \ce{H + trans\;HCSH -> CH2SH }
\item R12: \ce{H + CH3SH -> H2 + CH3S}          
\item R13: \ce{H + CH3SH -> H2 + CH2SH}         
\item R14: \ce{H + CH3SH -> H2S + CH3}  
\item R15: \ce{H + CH2SH2 -> H2 + CH2SH}
\end{itemize}

And for the radical-radical reactions:
\begin{itemize}
\item R16: \ce{H + HCS -> H2 + CS}              
\item R17: \ce{H + HCS -> H2CS}                         
\item R18: \ce{H + HCS -> trans\;HCSH}          
\item R19: \ce{H + HCS -> cis\;HCSH}            
\item R20: \ce{H + HSC -> H2 + CS}              
\item R21: \ce{H + HSC -> trans\;HCSH}          
\item R22: \ce{H + HSC -> cis\;HCSH}            
\item R23: \ce{H + CH3S -> H2 + H2CS}           
\item R24: \ce{H + CH3S -> CH3SH}               
\item R25: \ce{H + CH2SH -> H2 + H2CS}          
\item R26: \ce{H + CH2SH -> H2 + trans\;HCSH}   
\item R27: \ce{H + CH2SH -> H2 + cis\;HCSH}     
\item R28: \ce{H + CH2SH -> CH3SH}               
\item R29: \ce{H + CH2SH -> CH2SH2}
\end{itemize}

Table~\ref{benchmarkallbarrier} corresponds to Table~1 of the main manuscript but without zero-point energy corrections, likewise Table~\ref{benchmarkallbarrierless} corresponds to Table~2 of the main manuscript. For clarity, the following abbreviations are used: 
\begin{itemize}
 \item MPWB1K/def2-TZVP $\rightarrow$ MPWB1K,
 \item {B3LYP/def2-TZVP $\rightarrow$ B3LYP,}
 \item {PBEh-3c/mSVP $\rightarrow$ PBEh,}
 \item CCSD(T)-F12/VTZ-F12 $\rightarrow$ CC, 
\end{itemize}

In order to obtain an accurate rate constant, a correct description of the barrier region is important. Therefore, the discussion here will focus on calculations of the activation energy excluding zero-point energy corrections, $V_\text{act.}$. Values presented for the reaction energy, $V_\text{react.}$, are given for completion only. 

Although in general the DFT and CC values are in good agreement, with a maximal deviation of 2.3~kJ/mol, several T1 and D1 diagnostics are close to or surpass the typical thresholds, signalling the importance of taking multi-reference effects into account. Therefore, additional MRCI single point energies have been calculated. For the abstraction reactions from methyl mercaptan, even single point energies are too computationally expensive to calculate and T1 and D1 values are below the threshold in both cases. Generally, the MRCI values tend to correspond closely to those calculated with DFT and CC theory, {with a maximum deviation of 4.2~kJ/mol}. Furthermore, the contributions to the wave function of the various Slater determinants show that in all cases the main contribution consists of a Slater determinant with coefficient $ 0.911 - 0.954 $, showing that the multi-reference character is small. For comparison, the coefficient for the main contribution to the wave function for formaldehyde (\ce{H2CO}) is 0.943 when calculated at the MRCI(+Q)-F12/VTZ-F12 level.

\begin{table}[h]
 \centering\caption{Benchmark of DFT activation and reaction energies without zero-point energy contributions in kJ/mol against CC and MRCI single point energies for reactions with a barrier.}\label{benchmarkallbarrier}
 \begin{tabular}{l|rrr|rr}
 \hline
        & \multicolumn{3}{c}{$V_\text{act.}$}   & \multicolumn{2}{|c}{$V_\text{react.}$}   \\
                & {MPWB1K}       & CC    & MRCI  & {MPWB1K}               & CC              \\
\hline
  R1            & 2.7   & 1.8   & 2.3   & -247.8        & -230.8        \\
  R2            & 11.8  & 13.0  & 14.2  & -80.6         & -62.2         \\
  R3            & 35.6  & 37.9  & 39.8  & -38.1         & -37.9         \\
  R4            & 4.4   & 3.3   & 4.2   & -251.0        & -238.0        \\
  R5            & 1.8   & 0.7   & 1.5   & -208.3        & -195.7        \\
  R12           & 12.6  & 11.9  & --    & -67.8         & -72.2         \\
  R13           & 40.3  & 42.4  & --    & -25.1         & -30.0         \\
  R14           & 12.5  & 11.9  & --    & -60.1         & -69.1         \\
  \end{tabular}
\end{table}

\begin{table}[h]
 \centering\caption{Benchmark of DFT reaction energies without zero-point energy contributions  in kJ/mol against CC single point energies for reactions without a barrier.}\label{benchmarkallbarrierless}
\begin{tabular}{l|rrrr}
\hline
  Reaction      & \multicolumn{4}{c}{$V_\text{react.}$}         \\  
                & {MPWB1K}       & {B3LYP}  & {PBEh}  & CC           \\
\hline
  R16 & -199.5  & -211.5 & -194.2 & -228.4      \\
  R17 & -409.2  & -402.8 & -422.5 & -413.8      \\
  R18 & -219.8  & -215.0 & -205.4 & -224.0      \\
  R19 & -212.9  & -210.2 & -198.5 & -217.3      \\
  R20 & -366.7  & -379.2 & -378.4 & -397.0      \\
  R21 & -386.9  & -382.8 & -389.7 & -392.6      \\
  R22 & -380.0  & -378.0 & -382.7 & -385.9      \\
  R23 & -196.3  & -213.1 & -184.3 & -221.2      \\
  R24 & -379.4  & -372.5 & -371.3 & -386.9      \\
  R25 & -239.0  & -255.7 & -247.2 & -263.5      \\
  R26 & -49.5   & -68.0 & -30.1 & -73.6         \\
  R27 & -42.6   & -63.2 & -23.2 & -66.9         \\
  R28 & -422.2  & -415.1 & -434.2 & -429.2      \\
  \end{tabular}
  \end{table}

\section{Potential energy surface}\label{Profile}
\begin{figure*}
 \centering
 \includegraphics[width=18cm]{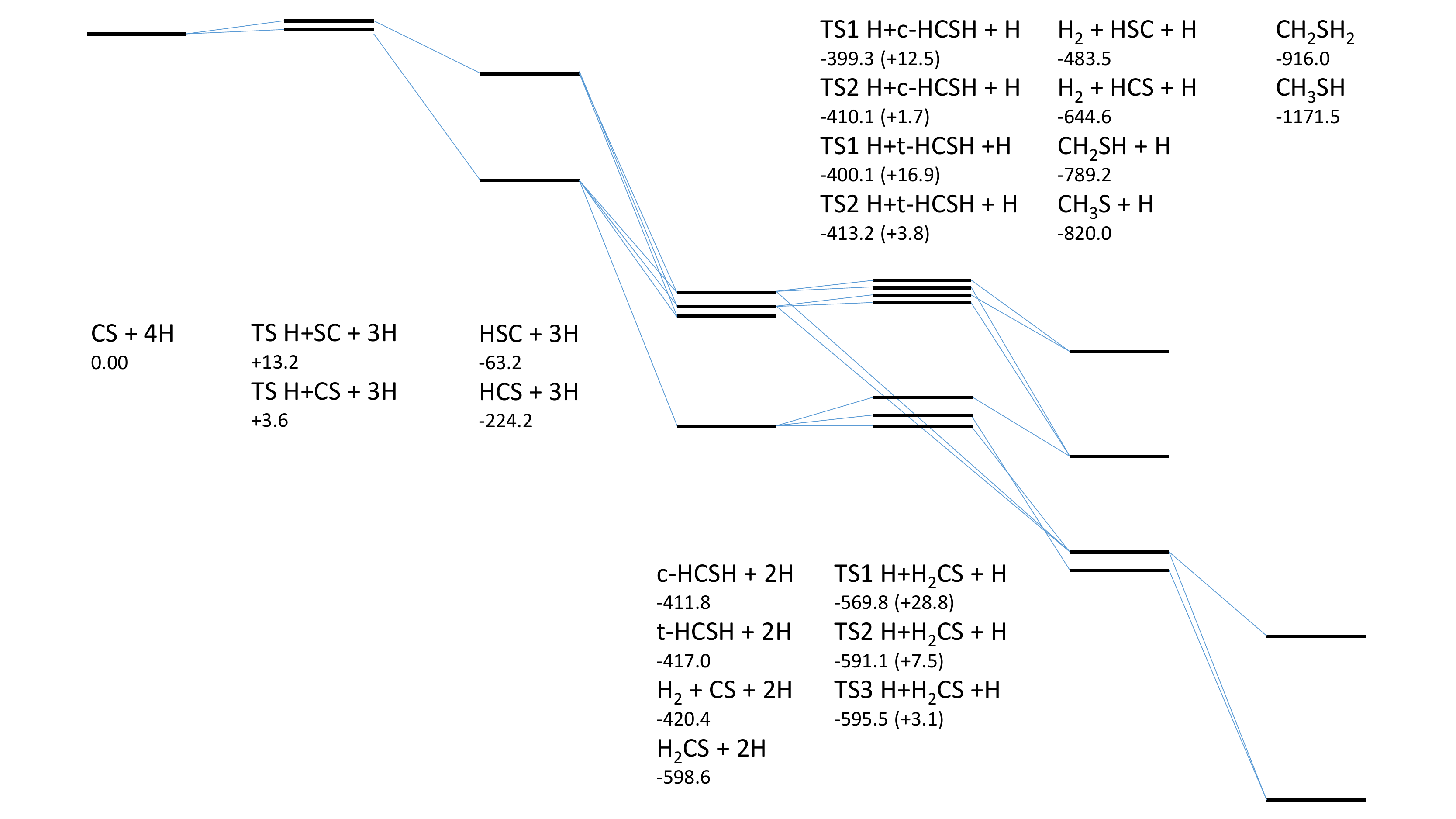} 
 \caption{Energy profile diagram for all species included in the network including the relative energies in kJ/mol.}\label{PES}
\end{figure*}

{Figure~\ref{PES} depicts the energy profile diagram for the species included in the reaction network, along with the transition states and relative energies. The barriers for the specific reactions are indicated in brackets.}

\section{Pre-reactive complex and transition state geometries}\label{Geom}

Both the pre-reactive complex (PRC) and transition state (TS) geometries for the reactions with a barrier in the gas phase and the reaction \ce{H + CS} on a water cluster are given in Tables~\ref{ClusterTS} and~\ref{TSH+CS}, calculated at the MPWB1K/def2-TZVP level of theory. 

\begin{table*}[h]
\centering
 \caption{TS geometries for reaction R1 \ce{H + CS} on a water cluster.}\label{ClusterTS}
 \begin{tabular}{lrrr|lrrr}
  \hline 
  \multicolumn{4}{l}{Cluster 1} &   \multicolumn{4}{l}{Cluster 2} \\     
  \hline
O &  -6.280716 &  -4.017202 &  -4.235239 & O &  -4.765945 & -3.121632 &  1.887968  \\
H &  -5.956881 &  -4.079146 &  -5.127344 & H &  -4.449591 & -2.995647 &  2.775852  \\
H &  -5.549508 &  -4.282264 &  -3.658088 & H &  -4.034988 & -3.493485 &  1.382132  \\
O &  -7.288372 &  -1.751707 &  -3.061103 & O &  -5.316986 & -1.137266 &  -0.039785 \\
H &  -8.155836 &  -1.962071 &  -2.731168 & H &  -4.651901 & -0.451231 &  0.117717  \\
H &  -7.005837 &  -2.522765 &  -3.578190 & H &  -5.288057 & -1.711944 &  0.734176  \\
O &  -0.536139 &  -1.983836 &  -1.046160 & O &  -2.185291 & -1.805626 &  -3.259418 \\
H &  -0.462832 &  -2.011162 &  -0.098791 & H &  -2.374169 & -1.072321 &  -3.840927 \\
H &  -1.231108 &  -1.346847 &  -1.248236 & H &  -3.047931 & -2.127300 &  -2.976451 \\
O &  -2.795405 &  -0.720380 &  -2.109602 & O &  -1.435912 & -1.308182 &  -0.603449 \\
H &  -3.635685 &  -1.066276 &  -1.755287 & H &  -1.727404 & -2.156339 &  -0.263866 \\
H &  -3.009013 &  0.112103  & -2.529093  & H &  -1.507957 & -1.396093 &  -1.563697 \\
O &  -4.247173 &  -4.511595 &  -2.429129 & O &  -4.527892 & -2.756707 &  -1.957331 \\
H &  -3.371433 &  -4.139223 &  -2.676029 & H &  -5.276025 & -3.138188 &  -2.405264 \\
H &  -4.092396 &  -5.365086 &  -2.038564 & H &  -4.882770 & -2.100278 &  -1.308600 \\
O &  -5.074581 &  -1.957475 &  -1.287975 & O &  -2.860240 & -3.841195 &  -0.064457 \\
H &  -5.871746 &  -1.769696 &  -1.797551 & H &  -3.442738 & -3.640021 &  -0.814579 \\
H &  -4.885023 &  -2.883807 &  -1.454455 & H &  -2.437852 & -4.672618 &  -0.255215 \\
O &  -2.007668 &  -3.249767 &  -3.094901 & O &  -3.187412 & 0.600096  & 0.158482   \\
H &  -1.304367 &  -3.226592 &  -2.439728 & H &  -2.849422 & 1.124943  & 0.875255   \\
H &  -2.307181 &  -2.333527 &  -3.094152 & H &  -2.482580 & -0.027782 &  -0.091807 \\
C &  -4.658415 &  1.280726  & -3.609752  & C &  -4.110602 & 0.664019  & -3.790438  \\
S &  -6.110849 &  1.245959  & -4.023575  & S &  -4.869326 & 1.346471  & -2.676164  \\
H &  -3.099714 &  1.002321  & -5.334113  & H &  -5.186039 & -1.111781 &  -4.878029 \\
  \hline
 \end{tabular}
\end{table*}

\begin{table*}[h!]\centering     
 \caption{PRC and TS geometries for the reactions in the gas phase.}\label{TSH+CS}
 \begin{tabular}{lrrr|lrrr}
\hline
  \multicolumn{4}{c}{PRC} & \multicolumn{4}{c}{TS} \\
 \hline
  \multicolumn{8}{c}{R1: \ce{H + CS -> HCS}}\\

H &    2.540730 & -0.042146 & -2.093731  & H   &  1.439531 &  0.006861 &  -1.366476 \\
C &   -0.563785 & -0.045342 &  0.206069  & C   & -0.369244 &  0.029111 &   0.174838 \\
S &    0.039358 & -0.011115 &  1.597700  & S   & -0.022173 & -0.033891 &   1.651435 \\
  \hline
  \multicolumn{8}{c}{R2: \ce{H + CS -> HSC}}\\
C &   0.137666 &  -0.001937 & -0.136981 & C &   0.138405 &  -0.001598 &  -0.122395 \\
S &  -0.240930 &   0.001533 & 1.331633  & S &  -0.189006 &   0.001486 &   1.366075 \\
H &   3.320907 &  -0.000648 & 3.962215  & H &   1.562623 &   0.000337 &   2.597380 \\
  \hline
   \multicolumn{8}{c}{R3: \ce{H + H2CS -> H2 + HCS}}\\  
H &  -3.423059 &  0.151238 & -2.814830  & H &   -1.665470 &   0.000493 &  -1.441435 \\
H &  -0.449577 &  0.003310 & -0.732812  & H &   -0.900503 &   0.002261 &  -0.802407 \\
H &   1.322700 & -0.160044 & -0.298113  & H &    1.041747 &  -0.001439 &  -0.585325 \\
C &   0.300471 & -0.064162 &  0.044881  & C &    0.118139 &   0.000341 &  -0.018332 \\
S &  -0.073988 & -0.025188 &  1.588018  & S &   -0.039352 &  -0.001662 &   1.537649 \\  
    \hline
  \multicolumn{8}{c}{R4: \ce{H + H2CS -> CH3S}}\\    
S &   0.662378 &  0.031094 & -0.004617  & S &   0.651757 &   0.038184 &  -0.004575 \\
C &  -0.880857 & -0.342694 & -0.000879  & C &  -0.912373 &  -0.282679 &  -0.000921 \\
H &  -1.443672 & -0.476508 & -0.915719  & H &  -1.470912 &  -0.430407 &  -0.915844 \\
H &  -1.438250 & -0.480563 & 0.916655   & H &  -1.466844 &  -0.429427 &   0.916645 \\
H &  -2.952116 &  2.968880 & 0.005245   & H &  -2.027033 &   1.680936 &   0.004493 \\
  \hline
  \multicolumn{8}{c}{R5: \ce{H + H2CS -> CH2SH}}\\    
C &   1.088978 &  0.131810 &  0.017982 & C  &  1.082200 &   0.138146 &   0.018062 \\
H &   1.578302 &  0.648658 & -0.797519 & H  &  1.568720 &   0.670758 &  -0.788671 \\
H &   1.718611 & -0.109749 &  0.864614 & H  &  1.713355 &  -0.123527 &   0.857289 \\
H &  -2.549387 &  2.715557 &  0.539801 & H  & -1.917367 &   1.893579 &   0.279935 \\
S &  -0.452431 & -0.247222 & -0.025240 & S  & -0.463426 &  -0.232647 &  -0.021670 \\  
  \hline
  \multicolumn{8}{c}{R12: \ce{H + CH3SH -> H2 + CH3S}}\\    
H &  -0.941393 & -0.619352 &  -0.504132 & H  & -0.938991 &  -0.616631 &  -0.511535 \\
H &   0.729513 & -0.146482 &  -0.770681 & H  &  0.741781 &  -0.160745 &  -0.741389 \\
H &  -0.478173 &  1.093225 &  -0.469581 & H  & -0.461374 &   1.093333 &  -0.482610 \\
C &  -0.172671 &  0.086313 &  -0.219730 & C  & -0.171981 &   0.086054 &  -0.215060 \\
S &   0.245912 & -0.061728 &   1.518349 & S  &  0.204014 &  -0.049595 &   1.530696 \\
H &  -0.951405 &  0.258743 &   2.005339 & H  & -1.072316 &   0.279311 &   1.988637 \\
H &  -3.878444 &  1.071088 &   2.947151 & H  & -2.267889 &   0.580063 &   2.207901 \\  
  \hline
  \multicolumn{8}{c}{R13: \ce{H + CH3SH -> H2 + CH2SH}}\\    
H &   3.451082 & -0.155284 & -1.251107 & H  &  1.808806 &   0.144998 &  -0.882658 \\
H &  -0.803721 & -0.841164 & -0.720646 & H  & -0.655960 &  -0.981294 &  -0.740924 \\
H &   0.584830 &  0.214229 & -0.508606 & H  &  0.906768 &   0.039124 &  -0.568357 \\
H &  -1.058852 &  0.883908 & -0.505498 & H  & -0.817415 &   0.798993 &  -0.583572 \\
C &  -0.441910 &  0.044440 & -0.214057 & C  & -0.342257 &  -0.095224 &  -0.208134 \\
S &  -0.575339 & -0.292140 &  1.545097 & S  & -0.489897 &  -0.349920 &   1.516783 \\
H &  -0.106775 &  0.878776 &  1.973043 & H  & -0.210823 &   0.895382 &   1.898157 \\
  \hline
  \multicolumn{8}{c}{R14: \ce{H + CH3SH -> H2S + CH3}}\\    
C &   -1.263211 & -0.027151 &   0.076904 &  C &  -1.277171 &  0.005097 &  0.045387 \\
H &   -1.543697 &  0.359316 &   1.047376 &  H &  -1.584492 &  0.398978 &  1.003842 \\
H &   -1.740330 &  0.545478 &  -0.706751 &  H &  -1.682026 &  0.603346 & -0.760577 \\
H &   -1.595755 & -1.054758 &   0.005485 &  H &  -1.627169 & -1.013394 & -0.053942 \\
S &    0.519886 & -0.062382 &  -0.131660 &  S &   0.547703 & -0.064821 & -0.088396 \\
H &    0.725013 &  1.249252 &  -0.028531 &  H &   0.732339 &  1.246579 &  0.003414 \\
H &    4.232205 & -0.220144 &   1.100397 &  H &   2.268859 & -0.142954 &  0.581416 \\   
\hline
  \end{tabular}
\end{table*}

\section{Unimolecular rate constants}\label{RC}

\begin{figure}[h]
 \centering
 \includegraphics[width=9cm]{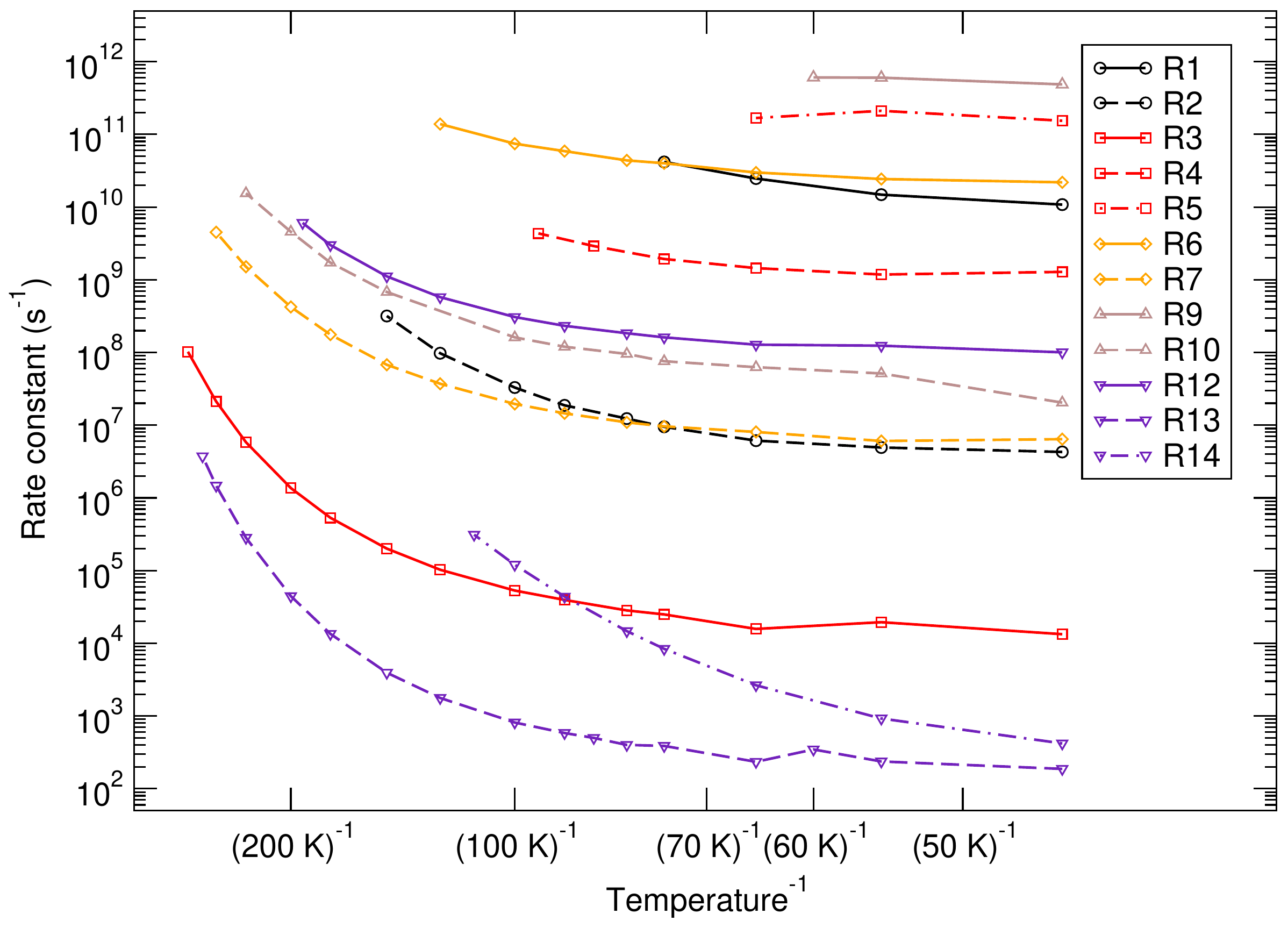}  
 \caption{Unimolecular reaction rate constants for reactions within the \ce{H + CS} network following the labelling indicated in Appendix~\ref{Bench}.}\label{RCFig}
\end{figure}

Values for the unimolecular rate constants are given in Table~\ref{RateConstants}.
\noindent The low-temperature values for the rate constant mentioned in Table~1 of the main manuscript are taken to be those at 45~K. This can be rationalized using Fig.~\ref{RCFig}. All but reactions~R10 and~R14 have roughly reached their asymptotic values, that is, the difference between the rate constant at 55~and 45~K being less than a factor 1.5. We note that the temperature dependence for the rate constant of reaction R14 behaves differently from the other reactions because a C-S bond breaking is involved there. As heavy-atom tunnelling is not efficient, the rate constant is consequently much lower than one might expect if only considering the activation energy. 

\begin{table*}[h!]\centering
 \caption{Temperature-dependent unimolecular rate constant values in s$^{-1}$.}\label{RateConstants}
 \begin{tabular}{rrrrrrr}
  \hline
  T (K) & R1 & R2 & R3 & R4 & R5 & R6\\
     \hline
45      & 1.08E+10      & 4.28E+06      & 1.34E+04      & 1.28E+09      & 1.54E+11 & 2.19E+10     \\
55      & 1.48E+10      & 4.92E+06      & 1.95E+04      & 1.18E+09      & 2.11E+11 & 2.43E+10     \\
65      & 2.47E+10      & 6.12E+06      & 1.58E+04      & 1.44E+09      & 1.68E+11 & 2.99E+10             \\
75      & 4.16E+10      & 9.48E+06      & 2.49E+04      & 1.93E+09      &          & 4.01E+10     \\
80      &               & 1.23E+07      & 2.83E+04      &               &          & 4.39E+10     \\
85      &               &               &               & 2.90E+09      &          &              \\
90      &               & 1.87E+07      & 3.95E+04      &               &          & 5.88E+10     \\
95      &               &               &               & 4.34E+09      &          &              \\
100     &               & 3.30E+07      & 5.31E+04      &               &          & 7.44E+10     \\
120     &               & 9.75E+07      & 1.03E+05      &               &          & 1.39E+11     \\
140     &               & 3.17E+08      & 2.00E+05      &               &          &              \\
170     &               &               & 5.32E+05      &               &          &              \\
200     &               &               & 1.37E+06      &               &          &              \\
250     &               &               & 5.81E+06      &               &          &              \\
300     &               &               & 2.12E+07      &               &          &              \\
370     &               &               & 1.02E+08      &               &          &              \\
                                                                                                 
  \hline                                                                                         
 T (K) & R7 & R9 & R10 & R12 & R13 & R14 \\
 \hline
45      & 6.42E+06      & 4.86E+11      & 2.05E+07      & 1.00E+08      & 1.87E+02 & 4.19E+02\\
55      & 6.06E+06      & 6.03E+11      & 5.15E+07      & 1.24E+08      & 2.37E+02 & 9.22E+02\\
60      &               & 6.07E+11      & 6.28E+07      &               & 3.46E+02 & \\
65      & 8.07E+06      &               &               & 1.28E+08      & 2.34E+02 & 2.65E+03\\
75      & 9.64E+06      &               & 7.57E+07      & 1.62E+08      & 3.88E+02 & 8.40E+03\\
80      & 1.09E+07      &               & 9.56E+07      & 1.84E+08      & 4.00E+02 & 1.47E+04\\
85      &               &               &               &               & 4.99E+02 & \\
90      & 1.45E+07      &               & 1.20E+08      & 2.33E+08      & 5.84E+02 & 4.38E+04\\
95      &               &               &               &               &          & \\
100     & 1.95E+07      &               & 1.61E+08      & 3.07E+08      & 8.15E+02 & 1.20E+05\\
120     & 3.72E+07      &               &               & 5.77E+08      & 1.77E+03 & 3.12E+05\\
140     & 6.77E+07      &               & 6.80E+08      & 1.12E+09      & 3.94E+03 & \\
170     & 1.76E+08      &               & 1.72E+09      & 3.01E+09      & 1.35E+04 & \\
190     &               &               &               & 6.05E+09      &          & \\
200     & 4.23E+08      &               & 4.57E+09      &               & 4.46E+04 & \\
250     & 1.51E+09      &               & 1.54E+10      &               & 2.83E+05 & \\
300     & 4.50E+09      &               &               &               & 1.47E+06 & \\
330     &               &               &               &               & 3.74E+06 & \\
  
  \hline
 \end{tabular}
\end{table*}

\end{document}